\documentclass[journal]{IEEEtran}

\ifCLASSINFOpdf
\else
   \usepackage{graphicx}
\fi
\usepackage{url}

\usepackage{graphicx}
\usepackage{float}
\usepackage{xcolor}
\usepackage{graphicx} 
\usepackage{epsfig}
\usepackage{epstopdf}
\usepackage{braket}
\usepackage[normalem]{ulem}
\usepackage[colorlinks]{hyperref}
\usepackage{siunitx}
\usepackage{soul}
\usepackage{mathtools}

\begin{document}

\title{\LARGE Single-Photon-Memory Measurement-Device-Independent Quantum Secure Direct Communication - Part \uppercase\expandafter{\romannumeral1}: Its Fundamentals and Evolution}

\author{Xiang-Jie Li, Dong Pan, \IEEEmembership{Member, IEEE}, Gui-Lu Long, \IEEEmembership{Member, IEEE}, and Lajos Hanzo, \IEEEmembership{Life Fellow, IEEE}

\thanks{Xiang-Jie Li is with State Key Laboratory of Low-dimensional Quantum Physics and Department of Physics, Tsinghua University, Beijing 100084, China.}

\thanks{Dong Pan is with Beijing Academy of Quantum Information Sciences, Beijing 100193, China.}

\thanks{Gui-Lu Long is with State Key Laboratory of Low-dimensional Quantum Physics and Department of Physics, Tsinghua University, Beijing 100084, China; Beijing Academy of Quantum Information Sciences, Beijing 100193, China; Frontier Science Center for Quantum Information, Beijing 100084, China; Beijing National Research Center for Information Science and Technology, Beijing 100084, China.}

\thanks{Lajos Hanzo is with School of Electronics and Computer Science, University of Southampton, Southampton SO17 1BJ, U.K.}

\thanks{Corresponding author's Email: Dong Pan, pandong@baqis.ac.cn; Gui-Lu Long, gllong@tsinghua.edu.cn.}

\thanks{This work is supported by the National Natural Science Foundation of China under Grants No. 11974205 and No. 12205011, the Key R\&D Program of Guangdong province (2018B030325002), Beijing Advanced Innovation Center for Future Chip (ICFC), Tsinghua University Initiative Scientific Research Program. }

\thanks{L. Hanzo would like to acknowledge the financial support of the Engineering and Physical Sciences Research Council projects EP/W016605/1 and EP/X01228X/1 as well as of the European Research Council's Advanced Fellow Grant QuantCom (Grant No. 789028)}
}

\maketitle

\begin{abstract}
   Quantum secure direct communication (QSDC) has attracted a lot of attention, which exploits deep-rooted quantum physical principles to guarantee unconditional security of communication in the face of eavesdropping. We first briefly review the fundamentals of QSDC, and then present its evolution, including its security proof, its performance improvement techniques, and practical implementation. Finally, we discuss the future directions of QSDC.
\end{abstract}


\IEEEpeerreviewmaketitle

\section{Introduction}

\IEEEPARstart{I}{nformation} security is vital to finance, national safety, corporate secrets and personal privacy. Traditional means of communication guarantee the reliable transmission of information over noisy channels, but it is unable to guarantee the unconditional security of transmitted information. Classical encryption is widely used for achieving secure transmission of information. However, due to the emergence of quantum computers, classical encryption is faced with severe challenges. For example, Shor's algorithm was shown to break both the Rivest-Shamir-Adleman (RSA) and other asymmetric encryption algorithms~\cite{shor1999polynomial}. Similarly, Grover's algorithm is capable of reducing the security of both the Advanced Encryption Standard (AES) and other symmetric encryption algorithms~\cite{grover1997quantum}. In order to cope with the security threats caused by quantum computing, researchers have improved the methods of key distribution, for example by using post-quantum cryptography~\cite{bernstein2017post}, which relies on specific mathematical problems that cannot be efficiently solved by quantum computers. Another design alternative is the quantum key distribution, which uses quantum states to negotiate secret keys~\cite{BB84}, and the secret keys will be used for secure classical communication.

\begin{figure}[H]
   \centering
    \includegraphics[width=\columnwidth]{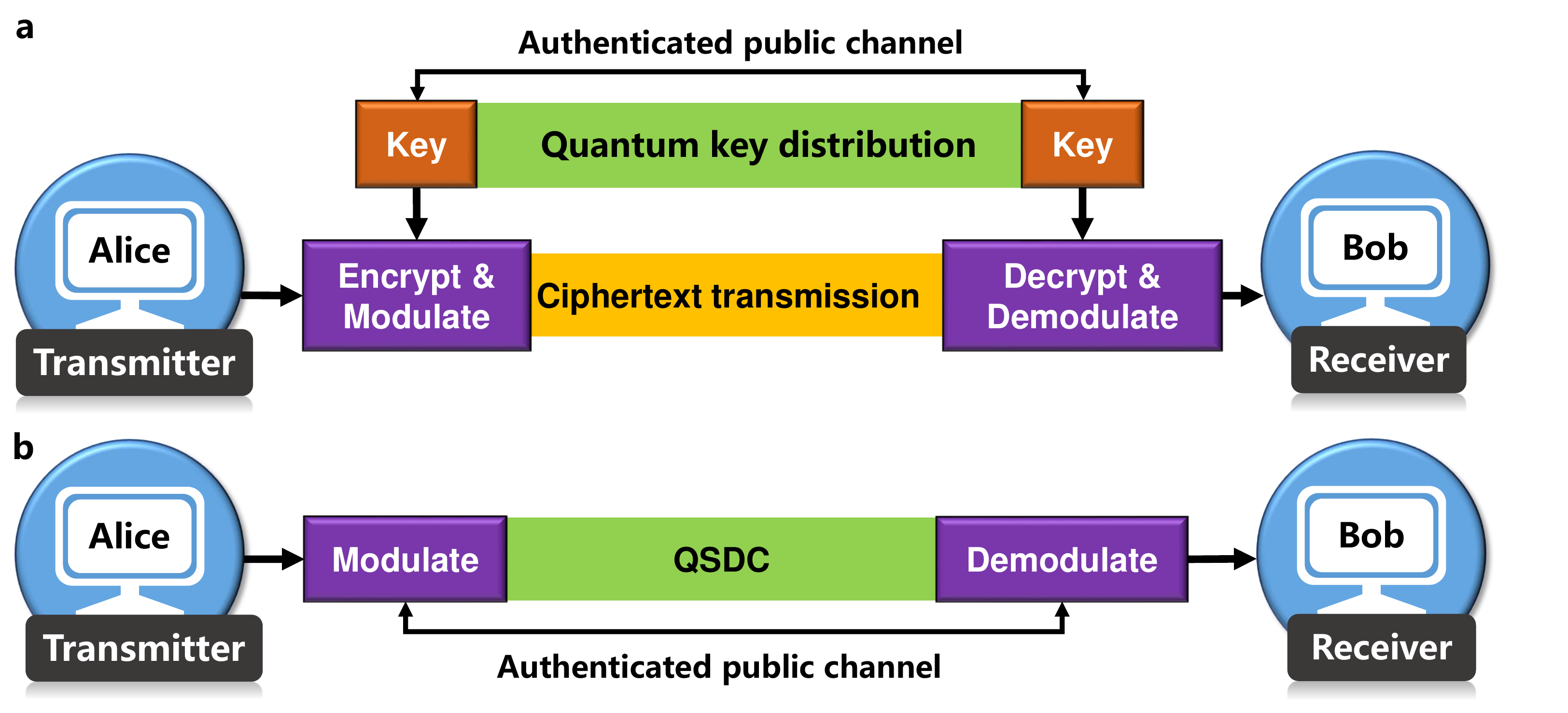}
    \caption{\label{fig:two-single}Secure communication frameworks. a: Twin-channel architecture. b: Single-channel architecture. QSDC: Quantum secure direct communication. Note that QKD uses an authenticated public channel for base sifting and post-processing, while QSDC uses this channel for eavesdropping detection and error correction.}
 \end{figure}

Quantum secure direct communication (QSDC) constitutes an attractive paradigm that transmits information directly using quantum states, which can guarantee both the reliability and security of the transmitted information at the same time~\cite{LL00}. Long and Liu proposed the first QSDC protocol in 2000~\cite{LL00}. It simplifies the twin-channel structure of Fig.~\ref{fig:two-single}a where ciphertext transmission and key distribution are separated, into a single-channel structure where only secret information is transmitted over the quantum channel, as illustrated in Fig.~\ref{fig:two-single}. This mitigated any potential security loopholes. Throughout the evolution of QSDC, a series of challenges have been encountered, as exemplified by the exact secrecy capacity calculation and security proof, the reliance on quantum memory, and the low performance due to the relatively weak quantum signal.

Against this background, in this letter, we survey the state of the art in QSDC. First, we briefly cover the implementation of QSDC by describing a basic QSDC protocol. Then we highlight a QSDC security analysis method based on quantum wiretap channel theory and apply it to QSDC protocols. Next, we introduce the key techniques for enhancing the performance of practical QSDC. Finally, we conclude this part by discussing the future directions and challenges faced by QSDC.

\section{\label{sec:level2}The fundamentals of QSDC\protect\\ 
}
QSDC relies either on entangled photon pairs or on single photons as quantum carriers. In experiments, entangled-photon pairs can be generated by parametric down-conversion~\cite{qi202115}, while single photons can be generated by attenuating laser pulses~\cite{hu16experimental}. In the following, we will describe the DL04 protocol~\cite{DL04} of Fig.~\ref{fig:DL04} relying on four steps:

\begin{figure}[H]
   \centering
    \includegraphics[width=\columnwidth]{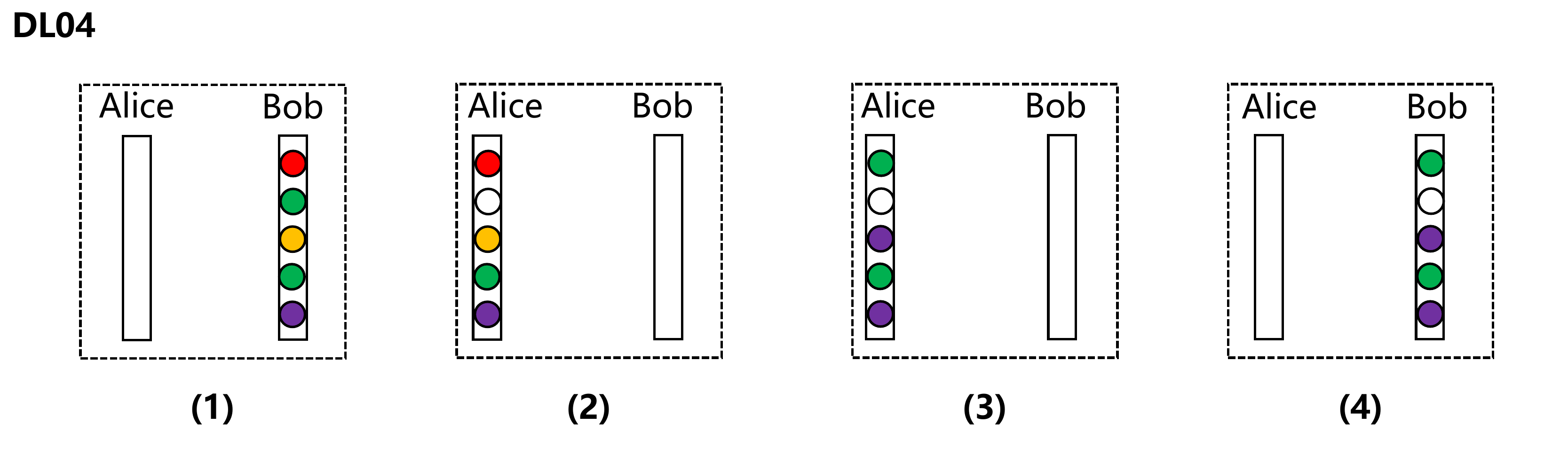}
    \caption{\label{fig:DL04}A basic protocol of QSDC: DL04 protocol.}
\end{figure}

\textbf{(1) Initialization:} Bob prepares $N$ single-photon states. \textbf{(2) Eavesdropping detection:} Bob sends single-photon state to Alice, and then Alice randomly selects some of the photons for measurement and informs Bob of the measurement results over an authenticated public channel of Fig.~\ref{fig:two-single}. If eavesdroppers contaminate the states, the measurement results of Alice will be different from the quantum states prepared by Bob. Alice and Bob can determine whether their link was contaminated by eavesdroppers based on the bit error rate. If there is no eavesdropper, the protocol continues. \textbf{(3) Information encoding:} Alice maps her bits 0 and 1 to quantum states, namely to the polarization or phase of the photons. \textbf{(4) Information transmission and integrity detection:} Alice sends the encoded quantum states to Bob, who measures them to infer the information. If the error rate is tolerable, the transmission is successful.
 
Note that Eve's action will perturb the quantum states and thus be detected by both communicating parties. Since QSDC performs eavesdropping detection before information encoding and transmission, this prevents Eve from obtaining any confidential information. The eavesdropping detection in Step (2) and the integrity detection in Step (4) ensure the secure and reliable transmission of information.

\section{\label{sec:level3}Security analysis\protect\\ }
\begin{figure}[H]
   \centering
    \includegraphics[width=\columnwidth]{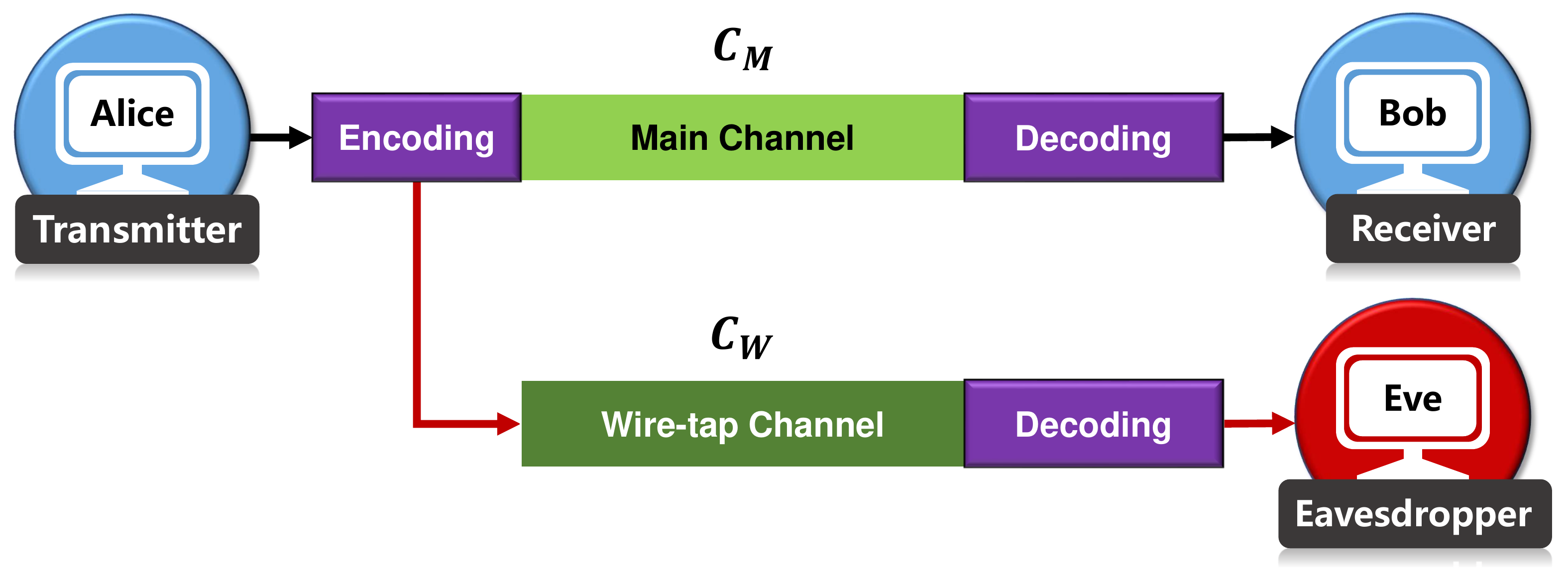}
    \caption{\label{fig:wiretap}The wiretap channel model. $C_M$ is the main channel capacity in the absence of Eve and $C_W$ is the wiretap channel capacity of Eve.}
\end{figure}

In order to accurately calculate the secrecy capacity, QSDC relies on the quantum wiretap channel model of~\cite{qi2019}, which is a generalization of Wyner's classical wiretap channel model of Fig.~\ref{fig:wiretap}. 
This theory demonstrates that a useful secrecy channel capacity $C_S$ above zero can be obtained, if the (main) channel capacity $C_M$ of the legitimate communication parties is higher than the eavesdropping (wiretap) channel capacity $C_W$, formulated as: 
\begin{equation}
   \label{Equ:Cs1}
   C_S=\max \limits_{\{p\}} \{I(A:B)-I(A:E)\}=C_M-C_W,
\end{equation}
where $p$ represents the probability of performing a unitary operation $U=I$ to encode bit 0, while $I(A:B)$ and $I(A:E)$ represent the mutual information between Alice and Bob as well as between Alice and Eve, respectively. Then there necessarily exists a coding method having a coding rate of $R \leq C_S$ that enables secure and reliable transmission of information. In the classical model, it is difficult to estimate Eve's wiretap channel capacity $C_W$, but QSDC allows us to calculate $C_W$ using the quantum bit error rate (QBER) estimate inferred from Eve's detection. 

Considering the security analysis of the DL04 protocol in~\cite{wu2019security} as an example, for the main channel capacity $C_M$, we assume that Alice and Bob communicate over a cascaded channel, namely a binary erasure channel and a binary symmetric channel. Then the QBER $e$ and communication reception rate of a quantum state $Q_{Bob}$ attained by eavesdropping detection in Step (4), allow us to write:
\begin{equation}
   C_M=Q_{\rm Bob}\cdot\left [1-h(e)\right ] ,
\end{equation}
where $h(\cdot )$ is the binary Shannonian entropy. 

For the wiretap channel capacity $C_W$, according to Eq.~(\ref{Equ:Cs1}), we have
\begin{equation}
   C_W=Q_{\rm Eve} \max I(A:E),
\end{equation}
where we denote Eve's reception rate of a quantum state by $Q_{\rm Eve}$. The maximum information Eve can infer is given by the Holevo bound
\begin{equation}
   \label{Equ:Holevo}
   \max I(A:E)=S(\sum_k p_k \rho_{\rm AE}^k)- \sum_k p_k S(\rho_{\rm AE}^k),
\end{equation}
where $S(\cdot )$ is the von Neumann entropy, $p_k=1/2$ and $\rho_{\rm AE}$ is the joint state of Alice and Eve given by
\begin{equation}
      \rho_{AE}=\mathnormal{\rm Tr_B}(\vert \Psi_{\rm ABE}\rangle \langle\Psi_{\rm ABE}\vert),
\end{equation}     
where 
\begin{equation}
   \vert \Psi_{\rm ABE} \rangle=\sum_{n = 1}^{4} \sqrt{\lambda _i}|\Phi _i \rangle |E_i\rangle,
\end{equation}    
is a quantum state shared by Alice, Bob and Eve~\cite{wu2019security}, while $|\Phi _i \rangle$ is a Bell state of system AB, and $\{|E_i\rangle\}$ is a set of orthogonal states of Eve's auxiliary system. The parameters $\lambda_i$ are constrained by the detected bit error rate (DBER) estimated by eavesdropping detection in Step (2), namely $\varepsilon _x=\lambda_2+\lambda_4$ and $\varepsilon _z=\lambda_3+\lambda_4$. After encoding, the state becomes 
\begin{eqnarray}
   \label{Equ:y}
   \begin{aligned}
        \rho_{\rm AE}^0&=U\cdot \rho_{\rm AE}\cdot U\\
        \rho_{\rm AE}^1&=Y\cdot \rho_{\rm AE}\cdot Y^{\dagger },
      \end{aligned}
\end{eqnarray}
where $U=I$, $Y=i\sigma_y$ are the quantum operations of Alice encoding the logical bit 0 and 1, respectively. We then obtain
\begin{equation}
   \label{Equ:Cw}
   C_W \leq Q_{\rm Eve}[h(\varepsilon _x+\varepsilon _z)].
\end{equation}
Finally, the secrecy capacity is formulated as
\begin{equation}
   \begin{aligned}
      C_S =& C_M-C_W\\
      \geq& Q_{\rm Bob}[1-h(e)]-Q_{\rm Eve}h(\varepsilon _x+\varepsilon _z).
   \end{aligned}
\end{equation}

\section{\label{sec:level4}Performance improvements\protect\\ }
During the development of QSDC, several challenges have been overcome. In order to improve the performance of QSDC, many innovative solutions have been developed~\cite{sun2020qmf,long2021drastic,zhou2020measurement,Chandra2022Direct,Lindsey2020Transmission,Shapiro:19}, as highlighted below.

\subsection{\label{subsec:QMF}Quantum-memory-free (QMF) QSDC}
In the original QSDC protocols introduced in Section~\ref{sec:level2}, Alice and Bob are required to select a fraction of the photons from $N$ photons for measurement in Step (2) of eavesdropping detection and retain the remaining photons for the subsequent operations. This process requires quantum memories. However, no high-performance quantum memory is available at the time of writing. Hence, there is a stumbling block in the practical implementation of QSDC. The above problem can be solved by using forward error correction coding (FEC)~\cite{sun2020qmf}.

\begin{figure*}
\centering
\includegraphics[width=.8\linewidth]{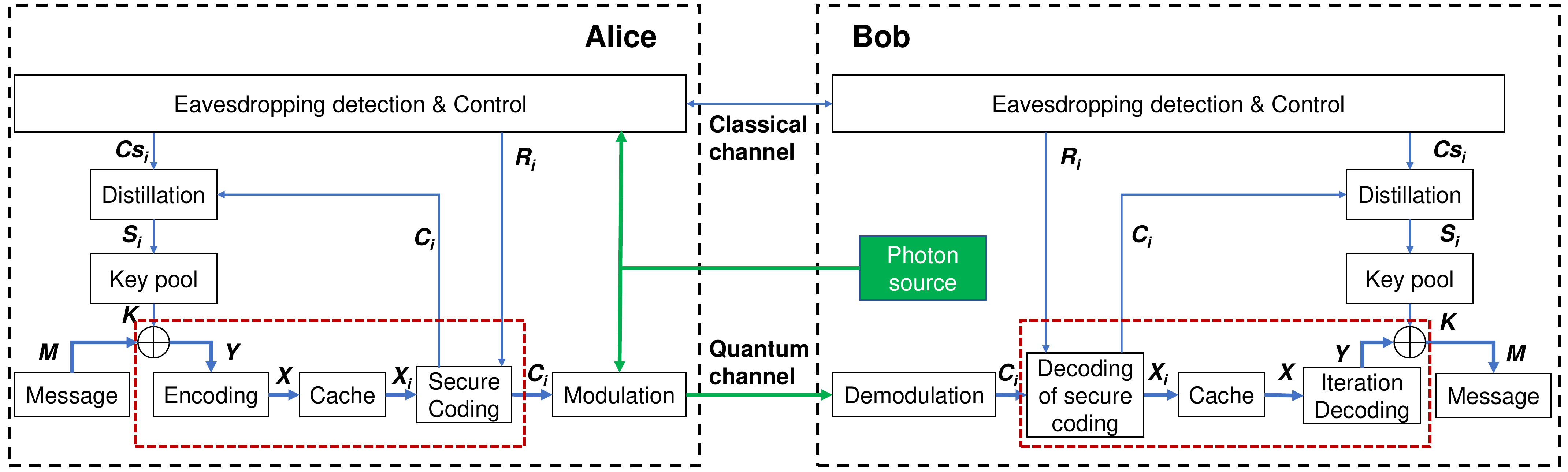}
\caption{\label{fig:QMF}Structure of the QMF-QSDC protocol and the data stream of the $i$-th frame~\cite{sun2020qmf}.}
\end{figure*}

A typical FEC-aided QSDC structure termed dynamic joint encryption and error-control coding is shown in the red dashed box of Fig.~\ref{fig:QMF}, where the FEC-coded information is divided into several frames for transmission, as detailed further below. The FEC precoding allows for reliable transmission of information by using for example powerful low-density parity-check (LDPC) codes. Secure coding relies on the idea of superimposing the secret key on the ciphertext, by arranging that each frame contains both the information of this frame and the key of the next frame. The information of each frame is mapped to quantum states. This regime allows simultaneous information detection and key distillation, thus ensuring that the next frame is secure. In this way, the need for quantum memory is removed.

The symbols in Fig.~\ref{fig:QMF} are defined as follows; $M\in \{0,1\}^m$ is the confidential information bit sequence; $K\in \{0,1\}^m$ is the key involved in encoding and decoding; $Y \in \{0,1\}^m$ represents the ciphertext which is the input of the FEC precoding module, described by $Y=M\oplus K$; $X \in \{0,1\}^k$ represents the codeword of the $(k,kR_p)$ LDPC code having a length of $k$, which is the output of the FEC precoding represented by the LDPC-encoding of the input sequence $Y$, where $m=kR_p$. The sequence $X_i\in\{0,1\}^{k_i}$ is either part of $X$ or constituted by a random bit sequence, which is the input of the secure coding module in the $i$-th frame, and $R_i$ is the rate of the secure coding in the $i$-th frame. Furthermore, $C_i\in \{0,1\}^{n_{c_i}}$ is transmitted to Bob over the quantum channel, which is a codeword of length $n_{c_i}$ produced by encoding $X_i$. After the transmission of $C_i$, Alice and Bob can infer the capacities $C_M$, $C_W$ and $C_S$ of Eq.~(\ref{Equ:Cs1}) in the system during the period in which the $i$-th frame is sent.

The above parameters are constrained by
\begin{equation}
\label{Equ:Cwi}
\frac{k_i}{n_{c_i}}\leq R_i- C_{W_{i-1}}, R_i < C_{m_{i-1}},
\end{equation}
which means that the secure coding rate of each frame has to be within the allowed range of the channel's secrecy capacity to ensure security. On the basis of Eq.~(\ref{Equ:Cw}) the wiretap channel capacity of the $(i-1)$-st frame is given by
\begin{equation}
C_{W_{i-1}}=Q_{\rm Eve}^{i-1}h(2\varepsilon ^{i-1} )=gQ_{\rm Bob}^{i-1}h(2\varepsilon ^{i-1} ),
\end{equation}
where $g=Q_{\rm Eve}/Q_{\rm Bob}$ and $\varepsilon ^{i-1}$ could be determined by experimental tests.

Let us now illustrate the specifics of the QMF-QSDC protocol. When the first round of communication is performed, the key pool of Fig.~\ref{fig:QMF} is empty. At this time, $X_i$ is composed of random numbers. Thus Alice and Bob can estimate $C_{M_1}$ and $C_{W_1}$ to distill the same key $S_1$ and put it into the key pool, which has to satisfy Eq.~(\ref{Equ:Cwi}). For the next action, there will now be enough keys in the pool, and the whole communication session consists of the following steps seen in Fig.~\ref{fig:QMF}: \textbf{(1)} Alice uses $K$ to encrypt $M$ to obtain $Y = M\oplus K$. \textbf{(2)} Encoding $Y$ into $X$, which is then stored in the cache. \textbf{(3)} Selecting $k_i$ bits from the cache and mapping them onto the quantum states, where $k_i$ and $R_i$ have to satisfy Equation~(\ref{Equ:Cwi}). \textbf{(4)} Sending the quantum states over the quantum channel to Bob. \textbf{(5)} Bob demodulates the received qubits and decodes the secure coding of Step 3. \textbf{(6)} After Bob receives $C_i$ correctly, both Alice and Bob can obtain $C_{M_i}$, $C_{W_i}$ and $C_{S_i}$. \textbf{(7)} If $C_{M_i}$ and $C_{W_i}$ satisfy Equation~(\ref{Equ:Cwi}), Alice and Bob are able to obtain the same new key $S_i$ by distilling $C_i$. Then we must repeat Steps (3) to (7) until all parts of $X$ are transmitted. \textbf{(8)} Finally, Bob uses the same $K$ to decrypt $Y$, by applying $M =Y\oplus K=M\oplus K\oplus K=M$. Note that if there are insufficient keys in the pool during the transmission, Alice and Bob have to transmit $X_i$ consisting of random numbers, which satisfy Eq.~(\ref{Equ:Cwi}) to distill a common key $S_i$.

The specific details of the coding algorithm which utilizes a generalized LDPC code based on Hadamard codes and repetition codes are not detailed in this compact Letter~\cite{sun2020qmf}.

\subsection{Increasing channel capacity using masking (INCUM) }
Eq.~(\ref{Equ:Cs1}) gives us the precise security capacity of QSDC, where $Q_{\rm Bob}$ and $Q_{\rm Eve}$ represent the appropriately varied reception rate of a quantum state for Bob and Eve. We generally assume that a powerful Eve is capable of using arbitrarily complex operations within the bounds of physical principles. Although the desired signals are degraded during transmission, Eve can collect them and achieves a higher reception rate than Bob. This reception rate difference increases, as the channel attenuation increases, which significantly degrades the security channel capacity of QSDC.

To improve the security channel capacity of QSDC, a simple yet powerful technique termed INCUM was proposed in~\cite{long2021drastic}. Consider the protocol in Section~\ref{sec:level2} as an illustration, where we can apply the INCUM method to Steps (3) and (4): \textbf{Step (3)} Alice generates a local random bit string $L$ and uses it to encrypt the message $M$, which will be transmitted to Bob for forming the ciphertext $M'$, namely $M'=L\oplus M$. Then she maps the logical bits 0 and 1 onto quantum states depending on $M'$. \textbf{Step (4)} Alice then sends the encoded quantum states to Bob, who measures them. Then Bob announces in which positions he has received qubits from Alice. Based on this announcement, Alice announces the random numbers used for encrypting photons in these positions. Finally, they perform security level checking. If the error rate is below the maximum tolerated value, the transmission is deemed successful.

By using the above method, Eve can not glean information from the lost quantum states, which makes the reception rate of Eve and Bob equal, namely $g=1$. This method only imposes low classical communication overhead, and yet it dramatically improves the secrecy channel capacity and transmission distance of QSDC.

\subsection{Measurement-device-independent (MDI) QSDC }
Although quantum communication has unconditional security in theory, when it comes to practical physical devices, security loopholes exist, since practical measurement devices may suffer from Trojan-horse attacks, fake states attacks, detector blinding attacks and so on. In order to resist these attacks, Measurement-device-independent QSDC protocols were developed, which guarantee that the communications still remain secure even if all measurement devices are controlled by an eavesdropper~\cite{zhou2020measurement}.

\begin{figure}[h]
\centering
\includegraphics[width=\columnwidth]{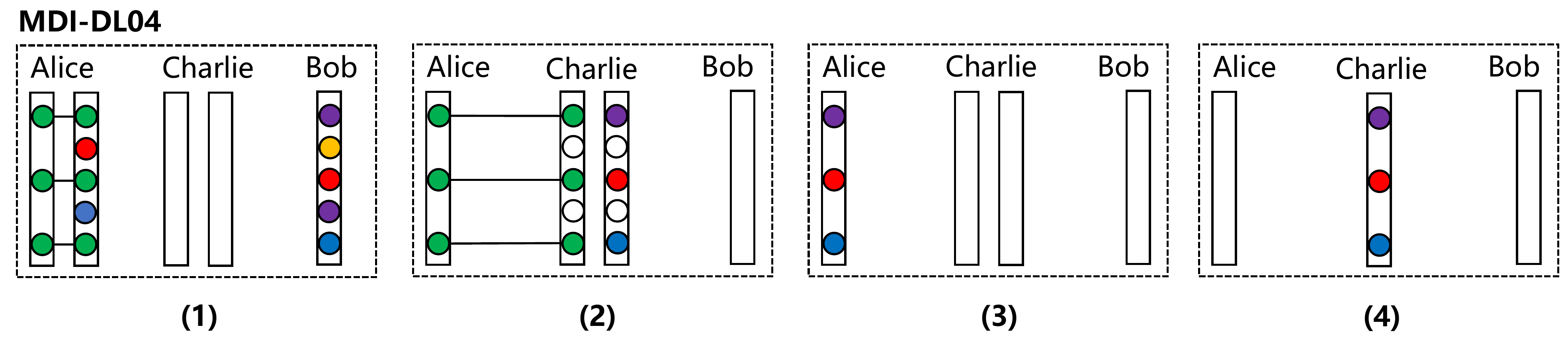}
\caption{\label{fig:MDI} The schematic of MDI-DL04 protocol.}
\end{figure}

The process of the MDI-DL04 protocol is shown in Fig.~\ref{fig:MDI}, where Alice and Bob send quantum states to an agent Charlie for measurement. The MDI protocols are innately secure, hence security is maintained even if the detector is completely controlled by the eavesdropper. Alice and Bob can determine whether Charlie is malicious based on the measurement results published by him. It obeys the following steps. \textbf{(1) Initialization:} Alice randomly prepares both entangled photon states and single-photon states, while Bob only prepares single-photon states. The entangled-photon states and the single-photon states are used for subsequent information transmission and eavesdropping detection, respectively. \textbf{(2) Eavesdropping detection:} Alice sends a single particle both in each entangled-photon pair and single-photon state to Charlie who performs Bell-state measurements~\cite{bennett1993teleporting}. If both Alice and Bob send a single-photon state, they can detect the existence of eavesdropping based on the measurement results. If Alice sends a single particle in her entangled-photon pair while Bob sends a single-photon state, they are said to perform quantum teleportation~\cite{zhou2020measurement}, in which Bob transmits his quantum state to Alice. \textbf{(3) Information encoding:} Alice then maps their logical bits to quantum states by employing quantum operations. \textbf{(4) Information transmission and integrity detection:} Finally, Alice sends her encoded quantum states to Charlie, who measures them. If the error rate is below the tolerable threshold, the MDI transmission is deemed successful.

Therefore, these MDI-QSDC protocols are designed for improving the security level of quantum communication.

\section{\label{sec:level5}Practical implementations\protect\\ }
QSDC has great application potential and economic value. To further improve its practicability, we report on a number of implementational breakthroughs.

\subsection{QSDC over 100 km fiber with time-bin and phase quantum states }
There have been several remarkable demonstrations of QSDC~\cite{hu16experimental,lum2016quantum,zhu2017experimental,zhang17experimental} in the last few years. The transmission distance of these experiments is between a few meters and ten kilometers. Very recently, Zhang \emph{et al}. designed a new experimental setup that achieved a QSDC distance of 100 km~\cite{zhang2022realization}. The new record attained by this experimental setup is mainly facilitated by sophisticated protocol modifications.

Observe from Eq.~(\ref{Equ:Cw}) that, the information leaked to Eve is bounded by $Q_{\rm Eve}[h(\varepsilon_x+\varepsilon_z )]$, where $\varepsilon_x$ and $\varepsilon_z$ are the DBER of the $X$-basis and $Z$-basis, respectively. The secrecy capacity is dependent on the error rate. However, Alice cannot derive a valid DBER using the $Y$-basis measurements, because using the $Y$-basis to measure the $X$-basis or $Z$-basis quantum state will lead to completely random results. Hence she has to infer the $Y$-basis DBER from the DBER of the $X$-basis and $Z$-basis. This results in a higher DBER than that calculated in Eq.~(\ref{Equ:Cw}). To obtain a higher secrecy rate, Zhang \emph{et al}. improved the protocol, which turned both the encoding basis and the detection basis into the $Z$-basis. The resultant wiretap channel capacity is given by~\cite{zhang2022realization}
\begin{equation}
\label{Equ:Cw2}
C_W \geq Q_{\rm Eve}[h(\varepsilon _z)].
\end{equation}
Based on Eq.~(\ref{Equ:Cs1}), a modest DBER reduction is capable of attaining a substantially higher increase in communication distance.

Zhang \emph{et al}.~\cite{zhang2022realization} replaced the FEC encoding scheme of~\cite{sun2020qmf} with a low-density BCH code, which was a concatenation of an LDPC code, a Bose-Chaudhuri-Hocquenghem code, and a repetition code for substantially increasing the secure communication capacity, distance, and rate.

\subsection{QSDC networks }
\begin{figure}[h]
   \centering
   \includegraphics[width=\columnwidth]{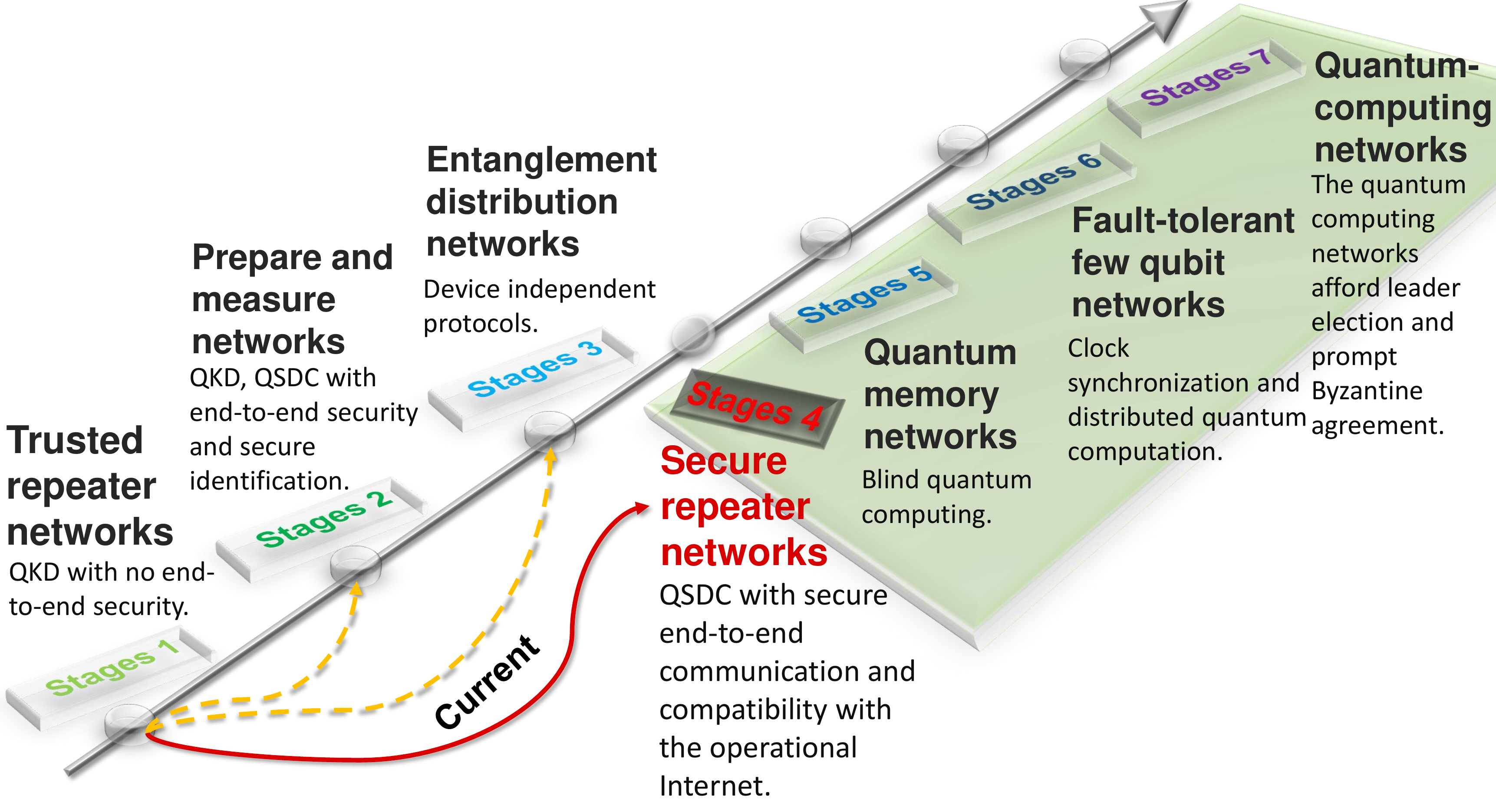}
   \caption{\label{fig:networks}Seven stages of quantum network development.}
\end{figure}

The ambitious goal of quantum communication is to establish a seamless global quantum network. However, the construction of the quantum Internet requires an evolutionary approach. As shown in Fig.~\ref{fig:networks}, at the time of writing a gradual transition is taking place from the trusted-repeater based networks to the prepare-and-measure networks and to entanglement-distribution based networks. Entanglement distribution QSDC networks were demonstrated in Ref.~\cite{qi202115}, while secure-repeater networks were reported in~\cite{long2022evolutionary}.

The trusted repeater networks require the repeater nodes to be absolutely trustworthy, which is difficult to guarantee in practical applications. Long \emph{et al}. proposed a secure repeater network~\cite{long2022evolutionary}, which solves the challenge of secure networking in quantum communication. In secure repeater networks, the users employ QSDC to transmit the ciphertext obtained by post-quantum cryptography, which can be securely relayed by a classical relay node. This supports end-to-end security in the networks and larger scale networks can be built in this way. These secure repeater networks can be implemented using current technology.
\section{Conclusion and future directions\protect\\}

In summary, we have presented the state of the art of QSDC, commencing from its fundamentals to its key technologies and practical implementations. QSDC supports the safe and reliable transmission of information. Some of the promising, future directions in QSDC are \textbf{(1)}. Given that QSDC is capable of 100 kilometres of practical optical fiber transmission, the next challenge is its free-space optical and quasi-optical/THz radio-frequency demonstration. \textbf{(2)}. Further QSDC research is required for supporting large-scale networking and the construction of the quantum Internet. \textbf{(3)}. Satellite-to-ground QSDC should also be investigated in the near future.


\end{document}